\documentclass[prd,preprint,superscriptaddress,showpacs,nofootinbib,tightenlines]{revtex4}
\usepackage{amsmath,amssymb,slashed,epsfig,subfigure,float,multirow}
\begin{document}
\title{Electromagnetic transition form factors of the Roper resonance in effective field theory}
\author{T.~Bauer}
\author{S.~Scherer}
\author{L.~Tiator}
\affiliation{PRISMA Cluster of Excellence, Institut f\"ur Kernphysik, Johannes Gutenberg-Universit\"at Mainz,
D-55099 Mainz, Germany}
\date{February 4, 2014}

\begin{abstract}
We analyze the form factors of the electromagnetic
nucleon-to-Roper-resonance transition in the framework of
low-energy effective field theory. A systematic power-counting
procedure is generated by applying the complex-mass scheme.
Within this power counting we calculate the form factors
to next-to-next-to-leading order and fit the results to
empirical data.
\end{abstract}

\maketitle
\section{Introduction}

In order to explore the structure of the nucleon and its excitations
a substantial experimental effort has been made to measure pion photo- and
electroproduction at electron accelerators such as Bates, ELSA, JLab,
and MAMI. Analyzing the available electroproduction
data, transition form factors for all four-star resonances below
center-of-mass energies of two GeV have been extracted in the
framework of phenomenological models (see, e.g., Refs.\
\cite{Tiator:2011pw,Aznauryan:2011qj,Aznauryan:2012ba,Crede:2013kia}
for an overview). Knowledge about the transition form
factors is necessary to obtain a complete understanding of the
nucleon excitation spectrum. In that context, precise data over a
wide range of momentum transfers have also been extracted for the
$P_{11}(1440)$ transition form factors
\cite{Aznauryan:2004jd,Drechsel:2007if,Aznauryan:2008pe,Tiator:2008kd,Aznauryan:2009mx,Mokeev:2012vsa}.
The $P_{11}(1440)$ resonance, often referred to as Roper resonance,
is the first excited state of the nucleon with quantum numbers
$I(J^P)=\frac{1}{2}(\frac{1}{2}^+)$ \cite{Roper:1964}. The
$Q^2$ dependence of the measured nucleon-to-Roper helicity
amplitudes supports the simple quark model assumption that it
constitutes the first radial excitation of the nucleon
\cite{Aznauryan:2009mx,Ramalho:2010mz}. On the other hand,
describing the Roper resonance in the framework of the simplest
spherically symmetric constituent quark model with SU$(6)$
spin-flavor symmetry leads to a parity reversal pattern since, in
contrast to the quark model calculations, the  $S_{11}(1535)$
turns out to be heavier than the Roper resonance
\cite{Isgur:1977ef,Isgur:1978wd}. By applying QCD-inspired
potentials, the right level ordering between the two resonances can
be generated within relativistic quark models \cite{Glozman:1995fu}.
Further understanding of the nature of the Roper resonance and the
level ordering is provided by lattice QCD. After also observing the
wrong parity reversal pattern in early studies
\cite{Sasaki:2001nf,Melnitchouk:2002eg,Edwards:2003cd}, recent
numerical simulations on the lattice hint towards the correct level
ordering
\cite{Lee:2002gn,Mathur:2003zf,Sasaki:2005ap,Mahbub:2009aa,Mahbub:2010me,Mahbub:2010rm}.

   Theoretical studies of the transverse and scalar (longitudinal)
helicity amplitudes have been performed in various frameworks such as
nonrelativistic constituent quark models (including relativistic corrections)
\cite{Copley:1969,Capstick:1992uc,Santopinto:2012nq}, relativistic quark models
 \cite{Weber:1990,Cardarelli:1996vn,Dong:1999cz,Ramalho:2010js},
chiral quark models \cite{Bermuth:1988ms,Alberto:2001fy,Golli:2009uk},
different hybrid models \cite{Li:1991yba,Obukhovsky:2011sc},
approaches including vector-meson-dominance features
\cite{Cano:1998wz,Vereshkov:2007xy},
and lattice QCD \cite{Lin:2008qv,Lin:2008gv}.
Even though the empirical data for the transverse and scalar (longitudinal)
helicity amplitudes can be described fairly well for larger values
of the squared momentum transfer in the framework of relativistic
quark models as well as in lattice QCD, neither of the two approaches
predicts the behavior in the low-$Q^2$ region correctly
\cite{Cardarelli:1996vn,Cano:1998wz,Alberto:2001fy,Aznauryan:2009mx,Ramalho:2010mz,Lin:2008qv,Lin:2008gv}.
The aim of this article is to investigate the electromagnetic
nucleon-to-Roper-resonance transition form factors in the framework of
low-energy effective field theory (EFT) which is based on chiral
perturbation theory \cite{Weinberg:1979kz,Gasser:1983yg} (see, e.g.,
Refs.\ \cite{Scherer:2002tk,Scherer:2012xha} for an introduction).
   Considering only the Goldstone boson sector of QCD, within this approach a
straightforward power counting, i.e., correspondence between the
loop expansion and the chiral expansion in terms of momenta and
quark masses at a fixed ratio \cite{Gasser:1983yg}, is obtained by
using dimensional regularization in combination with a minimal
subtraction scheme.
   The construction of a consistent power counting in effective field theories with
heavy degrees of freedom turns out to be a more complex problem
\cite{Gasser:1987rb} which can be resolved by choosing a suitable
renormalization scheme
\cite{Tang:1996ca,Becher:1999he,Gegelia:1999gf,Fuchs:2003qc,Schindler:2003xv}. In
order to also include resonant degrees of freedom, such as the Roper
resonance, we apply the complex-mass scheme (CMS)
\cite{Stuart:1990,Denner:1999gp,Denner:2006ic,Actis:2006rc,Actis:2008uh},
an extension of the on-mass-shell renormalization scheme to unstable
particles.
   In the context of the strong interaction, the CMS has successfully
been used to calculate the pole masses and widths of the $\rho$ meson \cite{Djukanovic:2009zn} and the
Roper resonance \cite{Djukanovic:2009gt}.
   Furthermore, electromagnetic properties have been investigated such as the magnetic
moments of the Roper resonance \cite{Bauer:2012at} and the $\rho$ meson \cite{Djukanovic:2013mka}
as well as the pion vector form factor in the timelike region \cite{Bauer:2011bv}.
   Finally, the CMS was shown to respect unitarity in a perturbative framework \cite{Bauer:2012gn}.

   This article is organized as follows.
   In Sec.\ II, we briefly discuss the effective Lagrangians on which the subsequent calculation is based.
   The applied renormalization scheme and the power-counting rules
are described in Sec.\ III.
   In Sec.\ IV, we give a definition of
   the electromagnetic transition form factors as well as the
   corresponding helicity amplitudes.
In Sec.\ V, we discuss the fit of our results to empirical data and analyze our final results.
   Section VI contains a short summary.

\section{Effective Lagrangian}

   In this section, we specify the effective Lagrangian
relevant for the subsequent calculation of the transition form factors
of the Roper resonance at next-to-next-to-leading order
(NNLO). Besides the pion, the nucleon, and the Roper resonance, we
also include the $\rho$ meson as an explicit degree of freedom. The
effects of other resonances such as the delta resonance are
parametrized in low-energy coupling constants. As is well known from
calculations of the electromagnetic form factors of the nucleon
\cite{Kubis:2000zd,Schindler:2005ke,Bauer:2012pv} and the
$\Delta$-resonance transition form factors \cite{Hilt}, the explicit
inclusion of the $\rho$ meson is essential for generating sufficient
curvature in the theoretically predicted results.

We write the effective Lagrangian
as\footnote{To simplify the notation, only bare masses are supplied
with a subscript $0$.}
\begin{equation}
\mathcal{L}_\text{EFT} =\mathcal{L}_{\pi}+ \mathcal{L}_{0}+
\mathcal{L}_{NR}+\mathcal{L}_{\rho}\,, \label{lagrFull}
\end{equation}
where $\mathcal{L}_{\pi}$ denotes the lowest-order Goldstone-boson
Lagrangian including the quark-mass term and the interaction with
the external electromagnetic four-vector potential ${\cal A}_\mu$
\cite{Scherer:2012xha}:
\begin{equation}
\label{l2} {\cal L}_\pi =
\frac{F^2}{4}\mbox{Tr}\left(\partial_\mu U \partial^\mu
U^\dagger\right) +\frac{F^2 M^2}{4}\mbox{Tr} \left( U^\dagger+ U
\right)+i\,\frac{F^2}{2}\mbox{Tr}\left[
\left(\partial^\mu U U^\dagger+\partial^\mu U^\dagger U\right)
v_\mu \right].
\end{equation}
   The pion fields are contained in the unimodular, unitary,
$(2\times 2)$ matrix $U$:
\begin{displaymath}
U(x)=u^2(x)=\exp\left(i\frac{\phi(x)}{F}\right),\quad
\phi=\phi_k\tau_k.
\end{displaymath}
   The external electromagnetic four-vector potential ${\cal A}_\mu$
enters into $v_\mu= - e\,{\cal A}_\mu \tau_3/2$
[$e^2/(4\pi)\approx 1/137, e>0$]. $F$ denotes the pion-decay
constant in the chiral limit, $F_\pi=F[1+O(\hat m)]=92.2$ MeV,
and $M$ is the pion mass at leading order in the quark-mass
expansion: $M^2=2 B\hat m$, where $B$ is related to the quark
condensate $\langle \bar q q\rangle_0$ in the chiral limit
\cite{Gasser:1983yg}.

   Introducing nucleon and Roper-resonance isospin doublets, $N$ and $R$, with bare
masses $m_{N 0}$ and $m_{R 0}$, respectively, $\mathcal{L}_{0}$
reads
\begin{equation}
\mathcal{L}_{0}
=\bar{N}\left(i\slashed{D}-m_{N0}+\frac{\texttt{g}_A}{2} \gamma^\mu\gamma_5 u_\mu \right)N
+\bar{R}\left(i\slashed{D}-m_{R 0}+\frac{g_R}{2}\gamma^\mu\gamma_5 u_\mu\right)R,
\label{lagL0}
\end{equation}
where $\texttt{g}_A$ corresponds to the chiral limit of the axial-vector coupling
constant, $g_A=1.2701(25)$ \cite{PDG}, and $g_R$ represents the analogue for the
Roper-resonance case.
   The building block $u_\mu$ is given by
\begin{equation}
u_\mu =i \left[u^\dagger \partial_\mu u -u \partial_\mu u^\dagger
-i\,\left( u^\dagger v_\mu u-u v_\mu u^\dagger\right)\right],
\label{umudef}
\end{equation}
and the covariant derivatives are defined as
\begin{equation}
\begin{split}
D_\mu H & =  \left( \partial_\mu + \Gamma_\mu-i\,v_\mu ^{(s)}\right) H, \\
\Gamma_\mu & =  \frac{1}{2}\,\left[u^\dagger \partial_\mu u +u
\partial_\mu u^\dagger-i\,\left( u^\dagger v_\mu u+u v_\mu u^\dagger
\right)\right], \label{cders}
\end{split}
\end{equation}
where $H$ stands for either the nucleon or the Roper resonance and
$v_\mu^{(s)}= - e{\cal A}_\mu/2$.
   By expanding $u_\mu$ of Eq.~(\ref{umudef}) for $v_\mu=0$ in terms of
the pion fields, one obtains from Eq.~(\ref{lagL0}) the Goldberger-Treiman relation
$\texttt{g}_{\pi NN}=m\, \texttt{g}_A/F$ \cite{Goldberger:1958,Nambu:1960},
where $\texttt{g}_{\pi NN}$ and $m$ denote the chiral limit of the
pion-nucleon coupling constant and the nucleon mass, respectively.
   An analogous relation results for the Roper resonance.

The interaction terms $\mathcal{L}_{NR}$ are constructed in
accordance with Ref.\ \cite{Borasoy:2006fk}. The leading-order
interaction between the nucleon and the Roper is given by
\begin{equation}
{\cal L}_{N R}^{(1)} = \frac{g_{N R}}{2}\,\bar R \gamma^\mu\gamma_5
u_\mu N+ {\rm H.c.},
\label{Rpiint11}
\end{equation}
where H.c.\ refers to the Hermitian conjugate and $g_{N R}$
is an unknown coupling constant.
The second- and
third-order Lagrangians for the nucleon-Roper-resonance interaction
relevant for our calculation read
\begin{equation}
\begin{split}
{\cal L}_{N R}^{(2)} & = \bar R\left[  \frac{c_{6}^*}{2}
\,f^+_{\mu\nu}+\frac{c_{7}^*}{2} \,v^{(s)}_{\mu\nu}
\right]\,\sigma^{\mu\nu}N+\text{H.c.}+\cdots \,,\\
{\cal L}_{N R}^{(3)} & = \frac{i}{2}\,d_{6}^* \bar R\left[
D^\mu,f^+_{\mu\nu}\right]\,D^\nu N+\text{H.c.}+ 2\,i\,d_{7}^* \bar
R\left(
\partial^\mu v^{(s)}_{\mu\nu}\right)D^\nu N+{\rm H.c.}+ \cdots \,,
\end{split}
\label{Rpiint2}
\end{equation}
where
\begin{equation}
\begin{split}
v_{\mu\nu}^{(s)} & = \partial_\mu v^{(s)}_\nu - \partial_\nu v^{(s)}_\mu\,,\\
f_{\mu\nu}^{+} & =  u f_{\mu\nu} u^\dagger +u^\dagger f_{\mu\nu} u\,,\\
f_{\mu\nu} & =  \partial_\mu v_\nu - \partial_\nu v_\mu-i
[v_\mu,v_\nu].
\end{split}
\label{bbks}
\end{equation}
   The coupling constants $c_{6}^*$ and $c_{7}^*$ are related to
the isovector and isoscalar transition magnetic moments.
   Furthermore, the coupling constants $d_{6}^*$ and $d_{7}^*$
contribute to the slopes of the isovector and isoscalar
transition form factors to be discussed below.
   As discussed in Ref.\ \cite{Borasoy:2006fk},
interaction terms of the form
\begin{align}
i \lambda_1 \bar R \slashed{D}N-\lambda_2 \bar R
N+\text{H.c.}
\label{lambdaiterms}
\end{align}
need not be considered.
   The first term and its Hermitian conjugate can be eliminated in terms of a
field transformation \cite{Scherer:1994wi} (equation-of-motion argument).
   After diagonalizing the nucleon-Roper mass matrix, the effects of the $\lambda_i$ terms
of Eq.\ (\ref{lambdaiterms}) can be absorbed in the couplings of already existing terms
or higher-order terms.

Finally, we need the Lagrangian containing the $\rho$ meson. The
$\rho$-meson triplet consists of a pair of charged fields,
$\rho_\mu^\pm=(\rho_{1\mu}\mp i\rho_{2\mu})/\sqrt{2}$, and a third
neutral field, $\rho_\mu^0=\rho_{3\mu}$. There are different
approaches to include vector mesons systematically into the
effective Lagrangian (see, e.g., Ref.\ \cite{Ecker:1989yg}). We
choose the $\rho$ meson to transform inhomogeneously under chiral
transformations $(V_L,V_R)$,
\begin{equation}
\rho_\mu\mapsto K\rho_\mu K^\dagger-\frac{i}{g}\partial_\mu K K^\dagger,
\label{transformation_rho}
\end{equation}
where
\begin{equation}
\begin{split}
\rho_\mu&=\rho_{k\mu}\frac{\tau_k}{2},\\
K(V_L,V_R,U)&={\sqrt{V_RUV_L^\dagger}}^{\,-1}V_R\sqrt{U},
\end{split}
\end{equation}
and $g$ is a coupling constant to be discussed below.
   The relevant part of the effective Lagrangian containing the $\rho$
meson can be written as
\begin{align}
{\cal L}_{\rho}={\cal L}_{\pi\rho}+{\cal L}_{\pi\rho N}+{\cal
L}_{\pi\rho R}+{\cal L}_{\pi\rho N R}.
\end{align}
The part describing the $\rho$ meson and its interaction with pions
reads \cite{Bauer:2012pv,Ecker:1989yg}
\begin{align}
\label{weinberg}
{\cal L}_{\rho}={}&-\frac{1}{2}\textrm{Tr}\left(\rho_{\mu\nu}\rho^{\mu\nu}\right)
+M_{\rho0}^2\textrm{Tr}\left[\left(\rho_{\mu}-\frac{i}{g}\Gamma_\mu\right)
\left(\rho^\mu-\frac{i}{g}\Gamma^\mu\right)\right]+\cdots\,,
\end{align}
where
\begin{align*}
\rho_{\mu\nu}&=\partial_\mu \rho_\nu-\partial_\nu
\rho_\mu-ig[\rho_\mu,\rho_\nu],
\end{align*}
and $M_{\rho0}$ denotes the (bare) $\rho$-meson mass.
   Note that the structure proportional to the low-energy constant (LEC) $d_x$ of Ref.\ \cite{Bauer:2012pv}
does not contribute to the {\it transition} form factors at NNLO and is, therefore, omitted from
Eq.~(\ref{weinberg}).
   The coupling constant $g$ can be fixed via the
Kawarabayashi-Suzuki-Riazuddin-Fayyazuddin relation
\cite{Kawarabayashi:1966kd,Riazuddin:1966sw},
\begin{equation}
M_\rho^2=2g^2F^2, \label{ksrf}
\end{equation}
generated by the combination of chiral symmetry and the consistency
of the EFT with respect to renormalizability
\cite{Djukanovic:2004mm}. Equation (\ref{weinberg}) is
self-consistent with respect to constraints and perturbative
renormalizability \cite{Djukanovic:2010tb}.

The remaining parts of the Lagrangian relevant for the subsequent
calculation are given by
\begin{equation}
\begin{split}
{\cal L}_{\pi\rho
N}={}&\bar{N}\left[k_1 \left(\rho_\mu-\frac{i}{g}\Gamma_\mu\right)\gamma^\mu\right]N+\cdots,\\
{\cal L}_{\pi\rho
R}={}&\bar{R}\left[k_2 \left(\rho_\mu-\frac{i}{g}\Gamma_\mu\right)\gamma^\mu\right]R+\cdots,\\
{\cal L}_{\pi\rho N
R}={}&\bar{R}\left[\frac{k_3}{2}\rho_{\mu\nu}\sigma^{\mu\nu}+k_4\left[D^\mu,\rho_{\mu\nu}\right]D^\nu\right]N
+\text{H.c.}+\cdots\,.
\label{roperrho}
\end{split}
\end{equation}
   In the following, we assume that the $\rho$ meson couples universally, meaning that the
self-coupling of the $\rho$ meson equals the coupling of the $\rho$ meson to pions and nucleons, $k_1=g$.
   Moreover, we also assume that the $\rho$ meson couples universally to the Roper
resonance, i.e., $k_2=g$.
   These universality conditions are supposed to be a consequence of
consistency conditions imposed by the demand of perturbative
renormalizability \cite{Djukanovic:2004mm}.

\section{Renormalization and power counting}

   In the following, we apply the CMS which originally was developed in the context of the Standard Model
   to derive properties of $W^\pm$, $Z^0$, and Higgs bosons obtained from
resonant processes
\cite{Stuart:1990,Denner:1999gp,Denner:2006ic,Actis:2006rc,Actis:2008uh}.
In Refs.\ \cite{Djukanovic:2009zn,Djukanovic:2009gt,Bauer:2012at,Djukanovic:2013mka},
the renormalization prescriptions were modified to obtain a
consistent power counting in the framework of EFT. Referring to
these articles, we split the bare parameters (and fields) of the
Lagrangian into, in general, complex renormalized parameters and
counter terms. We choose the renormalized masses as the poles of the
dressed propagators in the chiral limit:
\begin{equation}
\begin{split}
m_{R 0} &=  z_\chi+\delta z_\chi\,,\\
m_{N 0} &=  m+\delta m \,,\\
M_{\rho 0} &=  M_{\rho\chi}+\delta M_{\rho\chi} \,,
\end{split}
\label{barerensplit}
\end{equation}
where $z_{\chi}$ is the complex pole of the Roper propagator in the
chiral limit, $m$ is the mass of the nucleon in the chiral
limit, and $M_{\rho\chi}$ is the complex pole of the $\rho$-meson
propagator in the chiral limit. We include the renormalized
parameters $z_\chi$, $m$, and $M_{\rho\chi}$ in the propagators and
treat the counter terms perturbatively. The renormalized couplings
are chosen such that the corresponding counter terms exactly cancel
the power-counting-violating parts of the loop diagrams.

Since the starting point is a Hermitian Lagrangian in terms of bare
parameters and fields unitarity cannot be violated in the complete
theory. Generalizing the notion of cutting rules
\cite{Cutkosky:1960sp} to unstable particles, in Ref.\ \cite{Bauer:2012gn}
it was shown at the one-loop level that within the CMS unitarity
is also satisfied in a perturbative sense. In this context, it has
been verified that unstable particles do not appear as asymptotic
states and are therefore excluded from the unitarity condition
\cite{Veltman:1963th}.

We organize our perturbative calculation by applying the standard
power counting of Refs.\ \cite{Weinberg:1991um,Ecker:1995gg} to the
renormalized diagrams, i.e., an interaction vertex obtained from an
$O(q^n)$ Lagrangian counts as low-energy order $q^n$, a pion
propagator as order $q^{-2}$, a nucleon propagator as order
$q^{-1}$, and the integration of a loop as order $q^4$.
   In addition, we assign the order $q^{-1}$ to the Roper propagator
and the order $q^{0}$ to the $\rho$-meson propagator.
In practice, we implement this scheme by subtracting the loop
diagrams at, in general, complex ``on-mass-shell'' points in the
chiral limit. Since the virtual-photon four-momentum transfer $q^\mu$
counts as $O(q)$, we also assign the order $q^1$ to the mass difference between the Roper
resonance and the nucleon.

\section{Electromagnetic transition form factors}

The Roper resonance does not appear in the spectrum of asymptotic
states as it is an unstable particle. To define the matrix element
for the electromagnetic transition from the nucleon to the Roper
resonance, we consider the pion electroproduction amplitude for an
invariant energy near the mass of the Roper resonance. For the
incoming nucleon, being a stable particle, on-shell kinematics
correspond to $p_i^2=m^2_N$. On the other hand, for unstable
particles such as the outgoing Roper resonance, the analogous
kinematical point is given by the pole position, i.e., $p_f^2=z^2$.
  Introducing ''Dirac spinors'' $\bar w^i$ and $w^j$ with complex masses
$z$ for the final lines, in Ref.\ \cite{Gegelia:2009py}
  the authors described a method how to extract from the general vertex
only those pieces which survive at the pole.
   The renormalized vertex function for $p_i^2=m_N^2$ and $p_f^2=z^2$ may be
written in terms of two transition form factors,\footnote{The tilde symbol denotes
the phase convention of the present theoretical calculation.}
\begin{align}
 \sqrt{Z_R}\,\bar w^i(p_f) \Gamma^\mu(p_f,p_i)
u^j(p_i)\sqrt{Z_N}=\bar
w^{i}(p_f)\left[\left(\gamma^\mu-\slashed{q}\frac{q^\mu}{q^2}\right)\tilde{F}_1^*(Q^2)
+\frac{i\sigma^{\mu\nu}q_\nu}{M_R+m_N} \tilde{F}_2^*(Q^2)\right] u^j(p_i), \label{FF}
\end{align}
where $Z_R$ and $Z_N$ are the residues of the dressed propagators of
the Roper resonance and the nucleon, respectively. Moreover, we
introduced the positive-valued quantity $Q^2=-q^2=-(p_f-p_i)^2$. The
normalization of the Pauli form factor $\tilde{F}_2^*$ features the sum of
the nucleon mass $m_N$ and the Breit-Wigner mass of the Roper
resonance $M_R=1.44$ GeV. We choose this normalization to improve
comparability of our results with phenomenological analyses.
   Both transition form factors are complex-valued functions
even for $q^2<0$ because of the resonance character of the Roper.
  In contrast to the elastic case, the coefficient of the Dirac form
factor $\tilde{F}_1^*$ contains a term proportional to $q^\mu$
so that current conservation can be fulfilled.
   It is common to parametrize the nucleon-to-Roper transition
in terms of the transverse and scalar (longitudinal)
helicity amplitudes $A_{1/2}$ and $S_{1/2}$, respectively,
defined in the rest frame of the Roper resonance.
   The relation between the helicity amplitudes extracted from
experimental data and the matrix elements of the current operator
defines the coupling only up to a phase, which in the present
context reduces to a sign $\zeta=\pm 1$
\cite{Tiator:2011pw,Aznauryan:2012ba,Aznauryan:2007ja}.
   We therefore define
\begin{equation}
\label{eq:phase}
F_i^\ast(Q^2)=\zeta\tilde{F}_i^\ast(Q^2).
\end{equation}
   With this convention, the transverse and scalar (longitudinal) helicity amplitudes can be
expressed as the following linear combinations of the form factors
$F_1^*$ and $F_2^*$ \cite{Tiator:2008kd}:
\begin{equation}
\label{eq:AtoF}
\begin{split}
A_{1/2}(Q^2)&=\frac{e\;Q_{-}}{\sqrt{4K m_N M_R}}\Big(F_1^*(Q^2)+F_2^*(Q^2)\Big),\\
S_{1/2}(Q^2)&=\frac{e\;Q_{-}}{\sqrt{8 K m_N
M_R}}\left(\frac{Q_{-}Q_{+}}{2 M_R}\right)\frac{M_R+m_N}{Q^2}
\left[F_1^*(Q^2)-\frac{Q^2}{(M_R+m_N)^2}F_2^*(Q^2)\right],
\end{split}
\end{equation}
with
\begin{align*}
K&=\frac{M_R^2-m_N^2}{2M_R},\\
Q_{\pm}&=\sqrt{(M_R\pm m_N)^2+Q^2}.
\end{align*}
   According to \cite{Tiator:2011pw,Aznauryan:2012ba,Aznauryan:2007ja},
in Eq.~(\ref{eq:phase}) we choose $\zeta=-1$ if $g_{\pi NN}$ and $g_{\pi N R}$
have the same sign and $\zeta=1$ for opposite signs.
   In the present context, this translates into comparing the signs of
$\texttt{g}_A$ and $g_{NR}$.

\begin{figure}
\epsfig{file=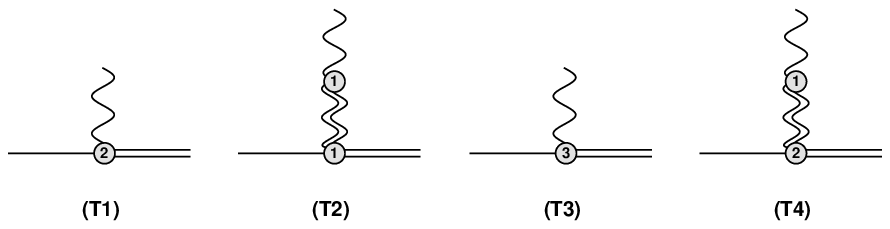,width=0.7\textwidth}
\caption[]{ \label{roperFFTree:fig} Tree diagrams contributing to
the electromagnetic transition form factors of the Roper resonance.
Solid and wiggly lines correspond to the nucleon and the
external electromagnetic field, respectively; double-solid lines
correspond to the Roper and double-wiggly lines to the $\rho$ meson.
The numbers in the vertices indicate the respective orders.}
\end{figure}

\begin{figure}
\epsfig{file=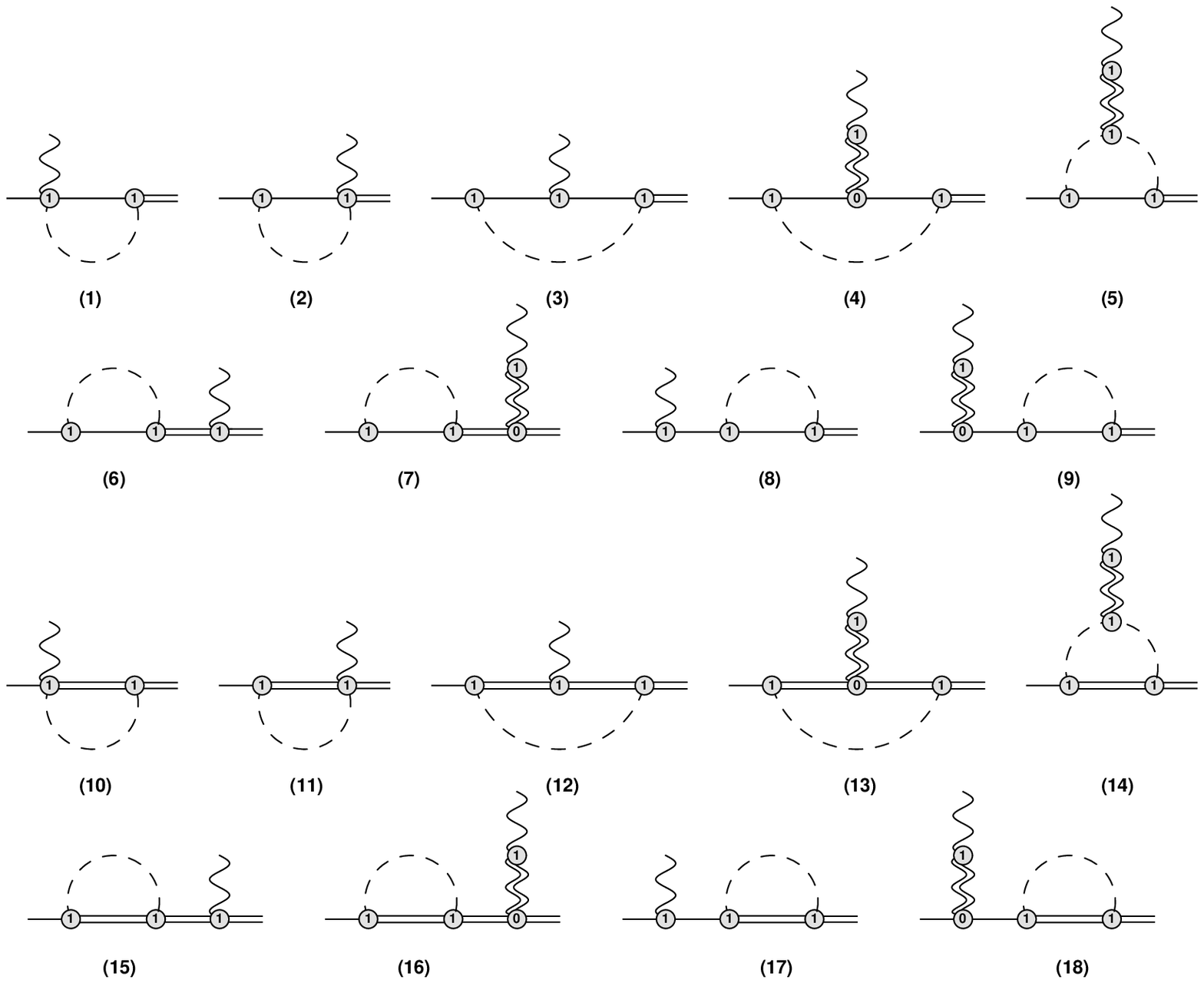,width=0.9\textwidth}
\caption[]{ \label{roperFFLoop:fig} Loop diagrams contributing to
the electromagnetic transition form factors of the Roper resonance.
Dashed, solid, and wiggly lines correspond to the pion,
nucleon, and external electromagnetic field, respectively;
double-solid lines correspond to the Roper and double-wiggly lines
to the $\rho$ meson. The numbers in the vertices indicate the
respective orders.}
\end{figure}

At NNLO [$O(q^3)$], the vertex function $\Gamma^\mu(p_f,p_i)$
obtains contributions from four tree diagrams (see Fig.\
\ref{roperFFTree:fig}) and eighteen loop diagrams (see Fig.\
\ref{roperFFLoop:fig}). According to Eq.\ (\ref{roperrho}), there is
no nucleon-photon-Roper and no nucleon-$\rho$-Roper interaction at
$O(q)$. Writing the wave function renormalization constant as
\begin{align}
Z_{N/R}=1+\delta Z_{N/R},
\end{align}
where $\delta Z_{N/R}$ is of $O(q^2)$, we find that the
product of tree diagrams (T1) and (T2) (see Fig.\
\ref{roperFFTree:fig}) and $\delta Z_{N/R}$ is at least of
$O(q^4)$, i.e., beyond the accuracy of our calculation.

To renormalize the diagrams of Fig.\ \ref{roperFFLoop:fig} we first
apply the modified minimal subtraction scheme of ChPT
($\widetilde{\text{MS}}$) \cite{Gasser:1987rb}. Then, we perform
additional finite subtractions such that the renormalized diagrams
satisfy the given power counting. We find that the
$\widetilde{\text{MS}}$-subtracted contributions to $F_1^*$ do not
contain any power-counting-violating terms. On the other hand, all
power-counting-violating terms of $F_2^*$ are analytic in small
quantities and can be absorbed by the renormalization of the available
coupling constants. This finding, together with current
conservation, provides an important consistency check for our
calculation.

\section{Numerical evaluation and results}

\renewcommand{\arraystretch}{1.4}
\begin{table}[t]
\caption{Physical values of the parameters, where the masses and
$F_\pi$ are given in units of GeV and $\texttt{g}_A$ is
dimensionless. The values are taken from Ref.\
\cite{PDG}. \label{parameter1}}
\begin{center}
\begin{tabular}{cccccc}
\hline
\hline
$m_N$& $M_\pi$& $z$ & $M_\rho$  & $F_\pi$ &  $\texttt{g}_A$ \\
\hline
$0.938$ \quad & $0.140$ \quad & $1.365-\frac{i}{2}\,0.190$ \quad & $0.775-\frac{i}{2}\,0.149$ \quad
 & $0.0922$ \quad & $1.27$ \\
\hline
\hline
\end{tabular}
\end{center}

\end{table}
\renewcommand{\arraystretch}{1}

For the numerical evaluation of the one-loop integrals we substitute
the physical values for the relevant parameters given in Table
\ref{parameter1}. The difference between the physical values and the
respective values in the chiral limit is beyond the given precision
with respect to a (chiral) low-energy expansion. More specifically,
we need to evaluate scalar one-, \mbox{two-,} and three-point functions with
complex parameters.\footnote{A definition of scalar loop integrals
can be found in Ref.\ \cite{Denner:2005nn}.} While it is well known
how to analytically continue scalar one- and two-point functions to
complex internal masses and complex external invariant momenta, the
evaluation of three-point functions with complex parameters is more
involved and, to the best of our knowledge, only possible for a few
special cases (see Ref.\ \cite{Passarino:2010qk}). In order to avoid
this complication, we drop the finite imaginary part of the internal
masses of the Roper resonance in all loop integrals.
Neglecting the imaginary part in the one-loop integrals
constitutes an effect of ${\mathcal O}(\hbar^2)$ which is beyond the
accuracy of our one-loop calculation.

   As far as the remaining eight LECs are concerned we make use of
the following strategy.
   The couplings $\texttt{g}_A$, $g_{NR}$, and $g_R$ appear in the
one-loop diagrams of Fig.~\ref{roperFFLoop:fig}, namely, in terms
of the products $\texttt{g}_A \cdot g_{NR}$ and  $g_R\cdot g_{NR}$ in
diagrams (1)-(9) and (10)-(18), respectively.
   In order to keep the number of free parameters as small as possible,
we follow Ref.~\cite{Borasoy:2006fk} and take $g_R=1$ such that
$g_A$ and $g_{R}$ are roughly of the same size (the naive quark model
predicts $g_A=g_R$).
   Furthermore, we explore the values of $g_{NR}$ in a region of $0.1-0.4$,
but use in our final result $g_{NR}=0.35$ \cite{Borasoy:2006fk}, compatible with
a fit to the branching ratio of the Roper resonance into $\pi N$.

   The remaining six parameters appear only in the tree diagrams and can
be determined by a fit to data.
   In total, four helicity amplitudes can be analyzed from electroproduction experiments,
tranverse ($_pA_{1/2}(Q^2),\;_nA_{1/2}(Q^2)$) and scalar (longitudinal) ($_pS_{1/2}(Q^2),\;_nS_{1/2}(Q^2)$)
helicity amplitudes of proton and neutron, respectively.
   From mainly JLab experiments measured by the CLAS collaboration, analyzed helicity amplitude data
are found in the literature from a MAID analysis, Ref.\ \cite{Tiator:2008kd}, and a CLAS analysis,
Ref.\ \cite{Aznauryan:2009mx}, both from single-pion electroproduction, and from a recent CLAS
analysis, Ref.\ \cite{Mokeev:2012vsa}, from two-pion electroproduction.
   These data points are only for a proton target and  $Q^2\geq 0.28$~GeV$^2$.
   No single-$Q^2$ data have yet been analyzed for the neutron target.
   At $Q^2=0$, the transverse helicity amplitudes are obtained from pion photoproduction
and rather precise values are found in the Particle Data Listings~\cite{PDG}.
   The scalar helicity amplitudes $_pS_{1/2}(0)$ and $_nS_{1/2}(0)$ are in general
also finite, but cannot be measured in a photoproduction experiment.

   The data points are obtained from reaction models describing the electroproduction
cross sections.
   MAID is a unitary isobar model incorporating all established resonances up to 2 GeV.
   The resonant contributions are parametrized in terms of a Breit-Wigner ansatz and the
background is given in terms of unitarized nucleon and vector-meson Born diagrams.
   A similar approach is used by the CLAS collaboration, including dispersion relations as an
additional constraint.
   The application of both models allows for the extraction of so-called single-$Q^2$
data for the electromagnetic transition from the proton to the positively charged Roper over a wide
range of momentum transfers.

   Because of the above-mentioned limitations in the data, we first performed a fit of the six
parameters to the empirical helicity amplitudes obtained in the analysis of MAID2007,
with an update in 2008, mainly due to new electroproduction data at higher $Q^2$~\cite{Tiator:2008kd}.
   This fit cannot describe all four helicity amplitudes simultaneously.
   In particular, we find different values for the coupling $k_4$,
which only affects the scalar helicity amplitudes, when fitted to proton and neutron amplitudes
separately.
   However, the empirical MAID fit that was found by a global fit of  the MAID model to
the world data base on electroproduction of proton and neutron targets, still has quite some
uncertainties, especially for the scalar neutron helicity amplitude, where practically no available
data was very sensitive to.
   As a consequence, we tried a fit to only three empirical helicity amplitudes,
by excluding the scalar neutron helicity amplitude.
   From this analysis with the empirical helicity amplitudes
we can conclude that a large isovector coupling of the $\rho$ meson is observed with a
value $k_3\approx -5.4$ together with a small $\pi NR$ coupling of $g_{NR}\approx 0.1$.

\begin{figure}[htbp]
 \centering
\includegraphics[width=1\textwidth]{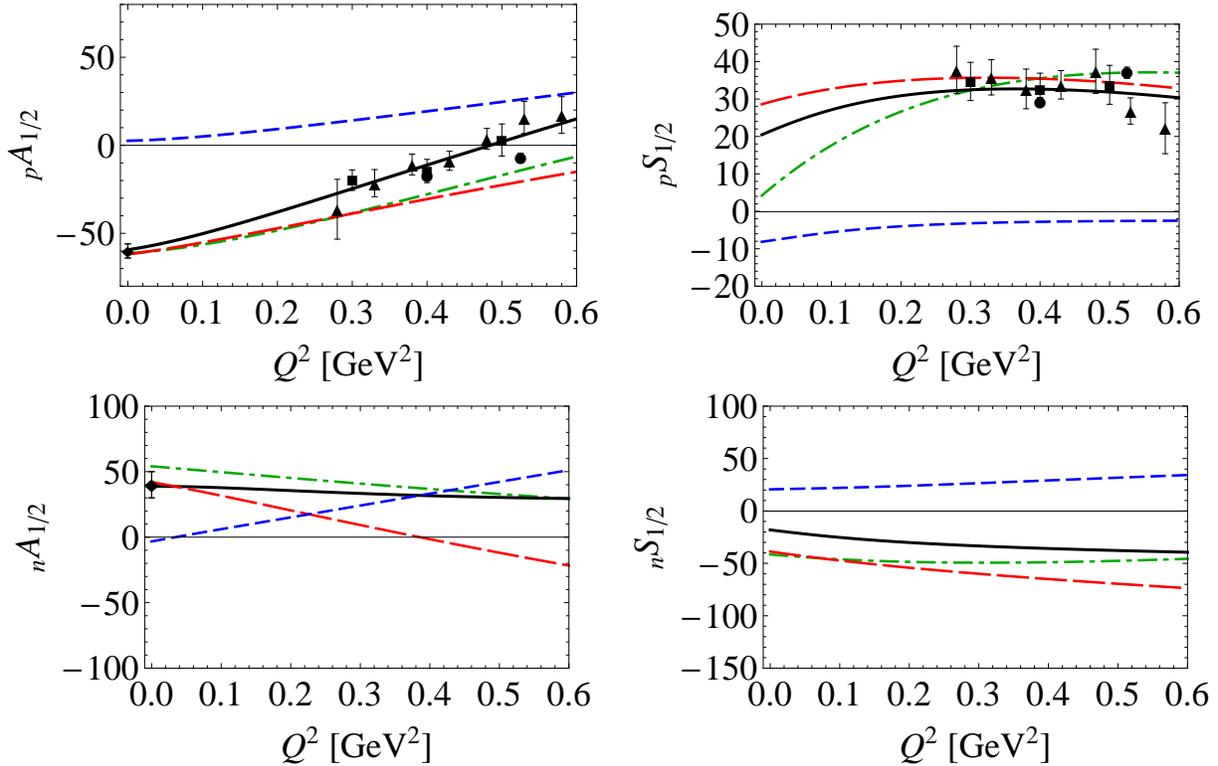}\\
\caption{(Color online) Transverse and scalar (longitudinal) helicity amplitudes
$A_{1/2}$ and $S_{1/2}$ of the
nucleon-to-Roper-resonance transition in units of $10^{-3}$ GeV$^{-1/2}$.
   Solid (black) lines: total results in the complex-mass scheme;
long-dashed (red) lines: tree contribution; short-dashed (blue) lines: loop contribution;
dash-dotted (green) lines: empirical parametrization of Ref.\ \cite{Tiator:2008kd}.
   The data points originate from the analysis of single-pion electroproduction
data (circles \cite{Drechsel:2007if} and squares \cite{Aznauryan:2009mx}) and double-pion
electroproduction data (triangles \cite{Mokeev:2012vsa}).
   The values of $A_{1/2}$ at the real-photon point are taken from Ref.~\cite{PDG}.
   The parameter values are for our fit 1, in particular, $g_{NR}=0.175$. }
\label{pRTransitionFFplot2}
\end{figure}

\begin{figure}[htbp]
 \centering
\includegraphics[width=1\textwidth]{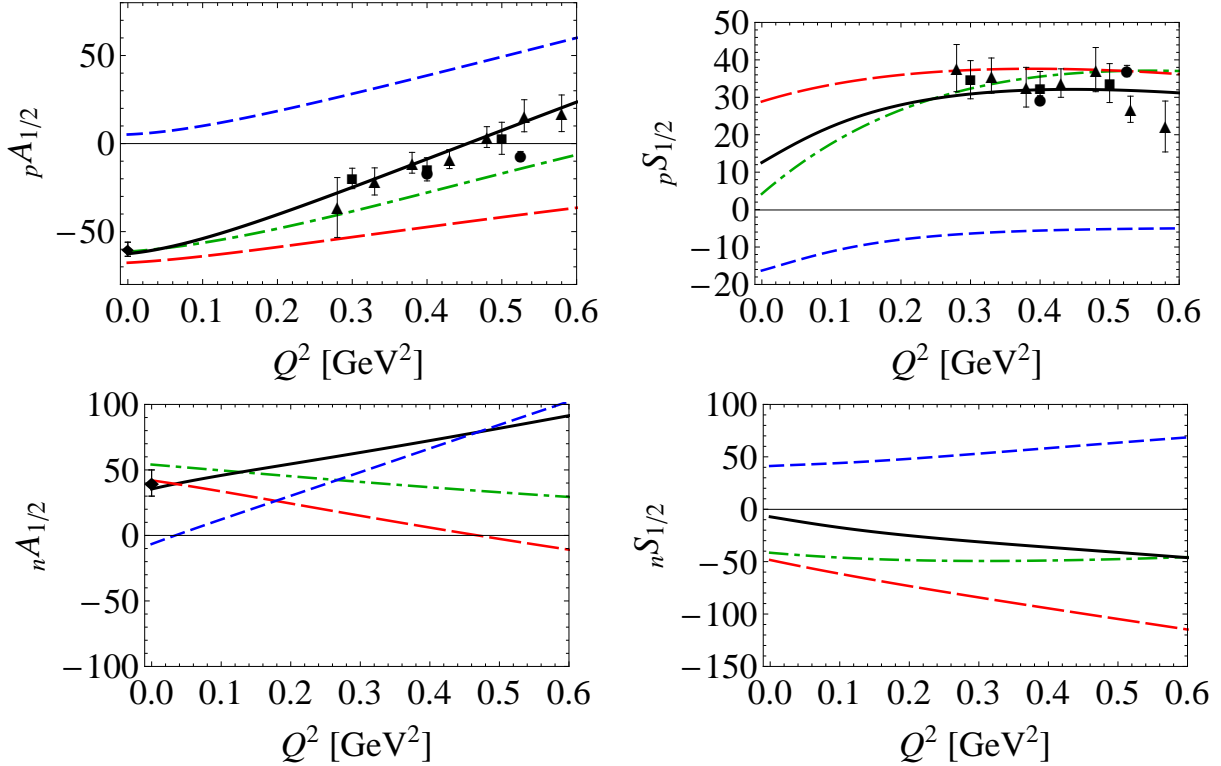}\\
\caption{(Color online) Same as in Fig.~\ref{pRTransitionFFplot2}, except that the parameter values
are for our fit 2, in particular, $g_{NR}=0.35$. } \label{pRTransitionFFplot}
\end{figure}

\begin{figure}[htbp]
 \centering
\includegraphics[width=1\textwidth]{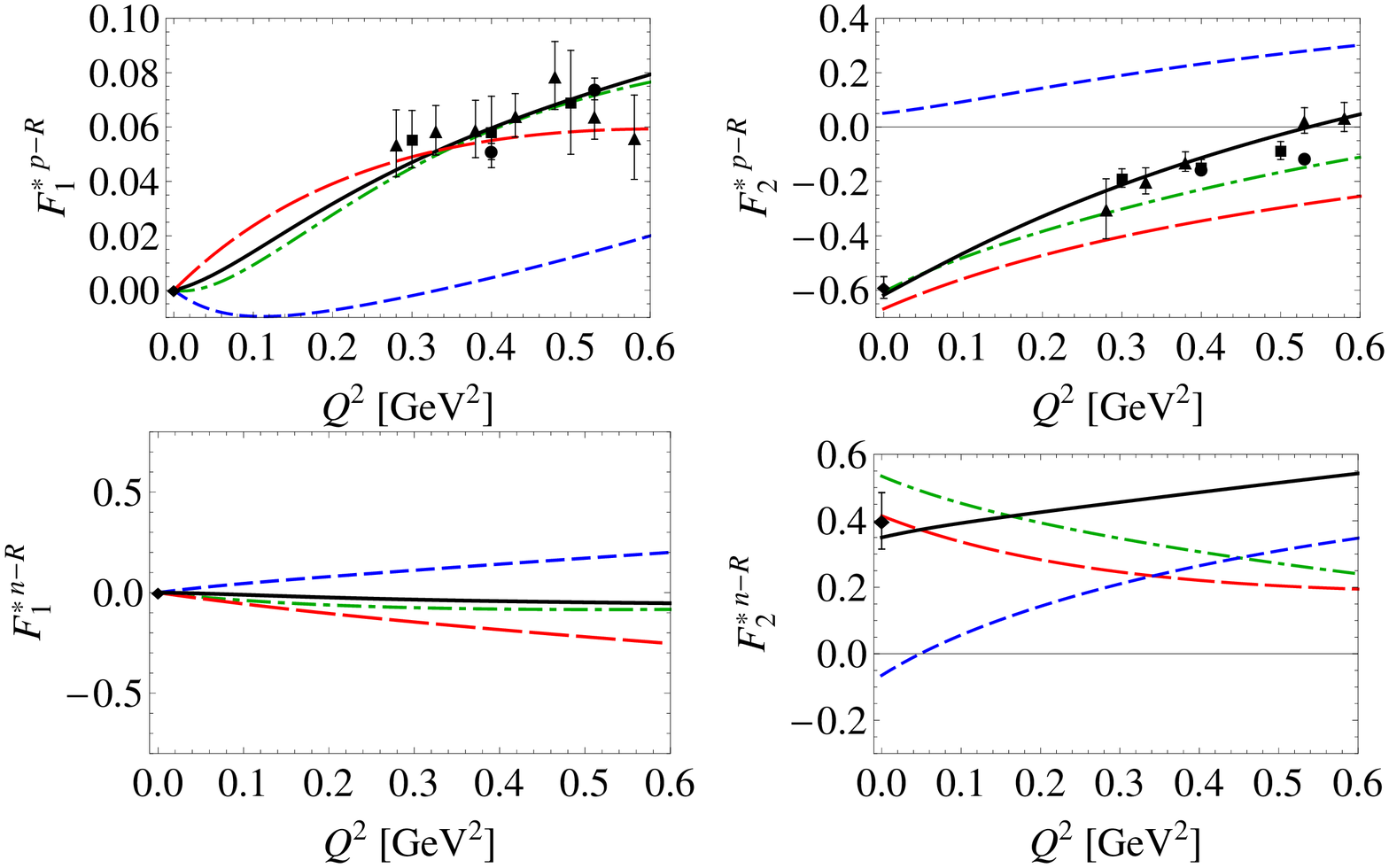}\\
\caption{(Color online) Dirac and Pauli form factors $F_1^*$ and $F_2^*$ of the nucleon-to-Roper-resonance
transition. The meaning of the curves and of the data points as in Fig.~\ref{pRTransitionFFplot2}.
The relations between the form factors and the helicity amplitudes is given in Eq.~(\ref{eq:AtoF}).
The parameter values are for our fit 2, in particular, $g_{NR}=0.35$. } \label{NRFFplot}
\end{figure}

   Comparing the empirical helicity amplitudes with data (see Fig.~\ref{pRTransitionFFplot2}),
a notable deviation is observed for the slope of the transverse proton helicity amplitude $_pA_{1/2}(Q^2)$.
   Therefore, we have refitted the $\gamma NR$ couplings for the proton $c_p^*, d_p^*$ to the available
single-$Q^2$ data.
   The results are shown in Fig.~\ref{pRTransitionFFplot2}, where we present our results together with
both the data and the empirical fit of the helicity amplitudes.
   In this first figure, the $\pi NR$ coupling was kept fixed at $g_{NR}=0.175$,
slightly larger than in the fit to the empirical helicity amplitudes, but only half of the value found in
Ref.~\cite{Borasoy:2006fk}.
   By increasing the $\pi NR$ coupling, the loop terms increase
accordingly and for the value of $g_{NR}=0.35$ we performed a second fit, see
Fig.~\ref{pRTransitionFFplot}.
   In a comparison with the data points, this second fit is very
similar, however, in comparison with the empirical scalar helicity amplitudes, larger deviations are
observed.
   Referring to the already discussed fact that the scalar helicity amplitudes
are based on very little experimental information, and the more realistic pion coupling, we
consider this second parametrization as our basic result.
  Our results for the LECs are shown in Table \ref{parameterTransitionFit}.
  In Fig.\ \ref{NRFFplot}, the resulting curves for the transition form
factors $F_1^\ast$ and $F_2^\ast$ are shown.

\renewcommand{\arraystretch}{1.4}
\begin{table}
\caption{Values for the fitted LECs, where $c_{p/n}^{*}=\frac{c_7^{*}}{2}\pm c_6^{*}$ and $k_3$ are
given in units of GeV$^{-1}$, $k_4$ in units of GeV$^{-2}$, and $d_{p/n}^{*}=d_7^{*}\pm d_6^{*}$ in
units of GeV$^{-3}$. The values of $g_{NR}$ in the second column were fixed for the two fits, see
text. \label{parameterTransitionFit}}
\begin{center}
\begin{tabular}{c c| c c c c c c }
\hline
\hline
$g_R$ & $g_{NR}$ & $c_{p}^{*}$ & $c_{n}^{*}$ & $d_{p}^{*}$& $d_{n}^{*}$ & $k_3$ & $k_4$\\
\hline
$1.0$ & $0.175$ & $0.27$ & $-0.18$ & $0.04$ & $0.04$ & $-5.5$ & $1.0$\\
$1.0$ & $0.35$ & $0.29$ & $-0.18$ & $0.04$ & $0.1$ & $-4.9$ & $1.0$\\
\hline\hline
\end{tabular}
\end{center}
\end{table}
\renewcommand{\arraystretch}{1.0}

\begin{figure}[htbp]
 \centering
\includegraphics[width=1\textwidth]{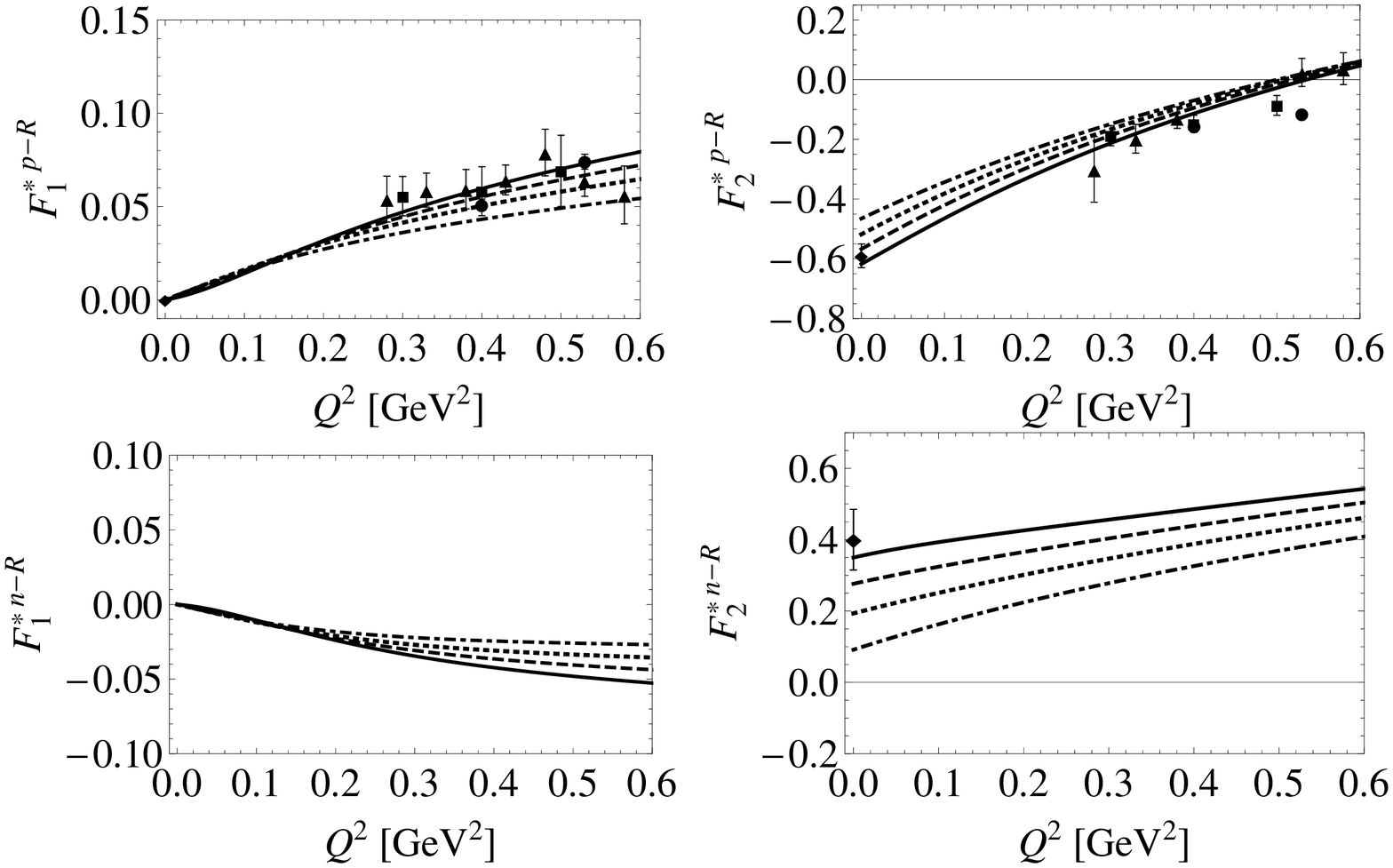}\\
 \caption{Dirac and Pauli form factors $F_1^*$ and $F_2^*$ of the nucleon-to-Roper-resonance
transition for different values of the pion mass. Solid lines refer
to $M_\pi=0.14$ GeV, dashed lines to $M_\pi=0.2$ GeV, dotted lines
to $M_\pi=0.3$ GeV, and dash-dotted lines to $M_\pi=0.4$
GeV, respectively.
The parameter values are for our fit 2, in particular, $g_{NR}=0.35$.}\label{Pionmass}
\end{figure}

   One possible approach to
obtain model-independent predictions for the transition form factors are
numerical simulations on the lattice.
   At present, calculations of the nucleon-to-Roper-resonance transition form
factors are based on the quenched approximation and seem to fail for
low squared momentum transfers \cite{Lin:2008qv,Lin:2008gv}. Given
the manifest Lorentz covariance of our results, they may
provide useful guidance for systematical extrapolations of lattice
simulations to the physical value of the pion mass. A fit of our
expressions to lattice data at different values for the pion mass
would result in a complete theoretical prediction of the transition
form factors and thereby a model-independent determination
of the LECs. To obtain an idea of the pion-mass dependence of the
transition form factors we show $F_1^*$ and $F_2^*$ for different
values of the pion mass in Fig.\ \ref{Pionmass}.

\section{Summary and outlook}
   To summarize, we have calculated the electromagnetic transition form factors
of the Roper resonance up to and including NNLO using EFT techniques.
   To obtain a systematic power counting, we applied the CMS which is a
generalization of the on-mass-shell renormalization to unstable particles.

   Our final results have been fitted to empirical data of the helicity amplitudes
up to and including $Q^2=0.58$ GeV$^2$.
   Even though the obtained results are in good agreement
with the empirical analyses, we stress that a one-loop calculation should
be treated with care beyond $Q^2=0.4$ GeV$^2$. The reason we have extended
the fits to such large values of $Q^2$ is the lack of empirical data in
the low-$Q^2$  domain.
   In order to reduce the number of fit parameters we fixed the LECs
$g_R$ and $g_{NR}$ since, in principle, they belong to other processes
and contribute only to loop diagrams in the present calculation.
   The remaining six LECs were assumed to be real and determined by
fitting the data of the proton helicity amplitudes and the empirical
parametrization of the neutron helicity amplitudes.
   A potentially useful application of our calculation is
in the context of lattice extrapolations. To that end, we have also
discussed the pion-mass dependence of the transition form factors.

In conclusion, it is possible to systematically calculate the
electromagnetic transition form factors of the Roper resonance
in the framework of
low-energy EFT applying the CMS and phenomenologically describe the
available empirical data in the low-$Q^2$ region. On the other hand,
due to the large difference between the nucleon and Roper resonance
masses, serving as an expansion parameter, the convergence of the
underlying perturbative expansion and the valid region of
applicability has yet to be studied more thoroughly.

The nucleon-to-Roper transition vertex containing the transition
form factors appears as a building block in the resonant $s$ channel
of pion electroproduction. For this reason, a full calculation of
pion electroproduction for center-of-mass energies in the region of
the Roper mass seems to be feasible choosing an appropriate power
counting.

\acknowledgments

When calculating the loop diagrams, we made use of the packages
FeynCalc \cite{Mertig:1990an} and LoopTools \cite{Hahn:1998yk}.
   The authors thank J.\ Gegelia for
  helpful
discussions and useful comments on the manuscript.
  This work was supported by the Deutsche Forschungsgemeinschaft
(SCHE459/4-1, SFB 443 and 1044).

\end{document}